\documentclass[aip,amsmath,amssymb,
reprint,%
]{revtex4-1}
\usepackage{color}
\usepackage{graphicx}
\usepackage{physics}
\usepackage{bm}
\usepackage{dcolumn}
\usepackage[utf8]{inputenc}
\usepackage[T1]{fontenc}
\usepackage{mathptmx}
\usepackage{gensymb}
\newcommand{\red}[1]{\color{black}#1}

\begin{document}

\title{ Dissipative Transport and Phonon Scattering Suppression via Valley Engineering in Single-Layer Antimonene and Arsenene Field-Effect Transistors}

\author{Jiang Cao}
\affiliation{
School of Electronic and Optical Engineering, Nanjing University of Science and Technology, Nanjing 210094, China
}
\email{jiang.cao@njust.edu.cn}
\author{Yu Wu}%
\author{Hao Zhang}
\affiliation{
Key Laboratory of Micro and Nano Photonic Structures (MOE) and Key Laboratory for Information Science of 
Electromagnetic Waves (MOE) and Department of Optical Science and Engineering, Fudan University, Shanghai 200433, China
}
\email{zhangh@fudan.edu.cn}
\author{Demetrio Logoteta}
\affiliation{
Dipartimento di Ingegneria dell’Informazione, Universit\`a di Pisa, Via G. Caruso 16, 56126 Pisa, Italy
}
\author{Shengli Zhang}
\affiliation{
MIIT Key Laboratory of Advanced Display Materials and Devices, College of Material Science and Engineering, Nanjing University of Science and Technology, Nanjing 210094, P. R. China
}
\email{zhangslvip@njust.edu.cn}
\author{Marco Pala}
\affiliation{ 
Universit\'e Paris-Saclay, Centre National de la Recherche Scientifique, Centre de Nanosciences et de
Nanotechnologies, 91120 Palaiseau, France
}

\date{\today}

\begin{abstract}
	Two-dimensional (2D) semiconductors are promising channel materials for next-generation field-effect transistors (FETs) thanks to their unique mechanical properties and enhanced electrostatic control. However, the performance of these devices can be strongly limited by the scattering processes between carriers and phonons, usually occurring at high rates in 2D materials. Here, we use quantum transport simulations calibrated on first-principle computations to report on dissipative transport in antimonene and arsenene $n$-type FETs at the scaling limit. We show that the widely-used approximations of either ballistic transport or simple acoustic deformation potential scattering result in large overestimation of the ON current, due to neglecting the dominant intervalley and optical phonon scattering processes. 
	We additionally investigate {\red valley engineering strategy  [Nano Lett. \textbf{19}, 3723 (2019)]} to improve the device performance by removing the valley degeneracy and suppressing most of the intervalley scattering channels via an uniaxial strain along the zigzag direction. 
	The method is applicable to other similar 2D semiconductors characterized by multivalley transport. 		  
\end{abstract}

\maketitle

\section{Introduction}

Recently, two-dimensional (2D) materials have shown a wealth of interesting properties and possible technological applications
 \cite{fiori2014a,novoselov2016a,schaibley2016a,yang2017a,jin2018a,liu2019a,das2019a,zhang2018a}.
Their mechanical properties, ultimate thinness and absence of surface dangling bonds are particularly suitable for conventional and flexible electronics \cite{fiori2014a,radisavljevic2011a,wang2012a,chhowalla2016a,cao2016a,pizzi2016a,polyushkin2020a}. 
The 2D materials with a finite energy bandgap 
have the potential to replace Si as the channel
material in ultra-scaled metal-oxide-semiconductor field-effect transistors (MOSFETs) for future nanoelectronics \cite{radisavljevic2011a,fiori2014a}.
However, two main issues hinder the adoption of 2D materials for MOSFET applications, namely the high resistance between metallic contacts  and semiconducting 2D materials \cite{allain2015electrical} and their relatively low mobility.

Experimental measurements of room temperature electron mobilities in 2D materials have so far reported values lower than 400~cm$^2$V$^{-1}$s$^{-1}$,~\cite{radisavljevic2011a,allain2014a,li2014a,liu2018a,sohier2018a} much smaller than the value for bulk Si (about 1400~cm$^2$V$^{-1}$s$^{-1}$)~\cite{cheng2020a,ponce2018a}.
When employed in electron devices in place of 3D semiconductors such as Si, Ge and III-V compounds, 2D-material-based MOSFETs do not suffer from surface roughness scattering~\cite{Poli_TED2008, Grillet_TED2017}, but have a room-temperature mobility limited by defects and electron-phonon coupling (EPC).
The former, just as the contact resistance, can in principle be reduced by developing dedicated engineering processes~\cite{C9NH00743A}. On the contrary, the electron-phonon scattering represents an intrinsic mechanism and cannot be easily suppressed at room temperature. 

Unfortunately, 2D semiconductors usually have larger electron-phonon scattering rates with respect to their 3D counterparts, due to a higher density of electronic and phononic states at energies close to the band extrema~\cite{cheng2020a}. 
The low phonon-limited mobility in monolayers has been also confirmed by several first-principle calculations based on density functional theory (DFT) and on the Boltzmann transport equation~\cite{kaasbjerg2012a,sohier2018a,liu2019a,fischetti2016a,sohier2016a,cheng2018a,cheng2019a}. 

In light of these considerations, it appears compulsory to take into account the electron-phonon interactions in the simulation of 2D material-based electron devices, in order to obtain reliable quantitative predictions of the device performance at room temperature. 
Nevertheless, only a few recent simulation studies have fully included the effect of EPC~\cite{szabo2015a,Logoteta_PhysRevRes2020}. In other works, the mobility has been evaluated through the Takagi formula~\cite{takagi1994a}, which takes  into account only the Bardeen-Shockley acoustic-deformation-potential (ADP) scattering~\cite{bardeen1950a}  and neglects all the optical and intervalley phonon scatterings~\cite{wang2017a,pizzi2016a}. However, many 2D semiconductors have multiple degenerate conduction band valleys, which results in a large phase space for intervalley scatterings and in large scattering rates~\cite{sohier2018a,sohier2019a}. The inelastic nature of these scattering processes has been already shown to deeply affect the transport in electronic devices~\cite{szabo2015a,pala2016a}.

More often, only the ballistic regime is considered in simulations. This approximation is usually justified by invoking the shortness of the considered device, though no experimental evidence of ballistic transport at room temperature for 2D materials other than graphene has been provided yet. Therefore, it is still unclear to which extent the electron-phonon scattering affects the room-temperature transport even in devices with characteristic lengths of less than ten nanometers.

\begin{figure}
	\centering
        	\includegraphics[width=\linewidth]{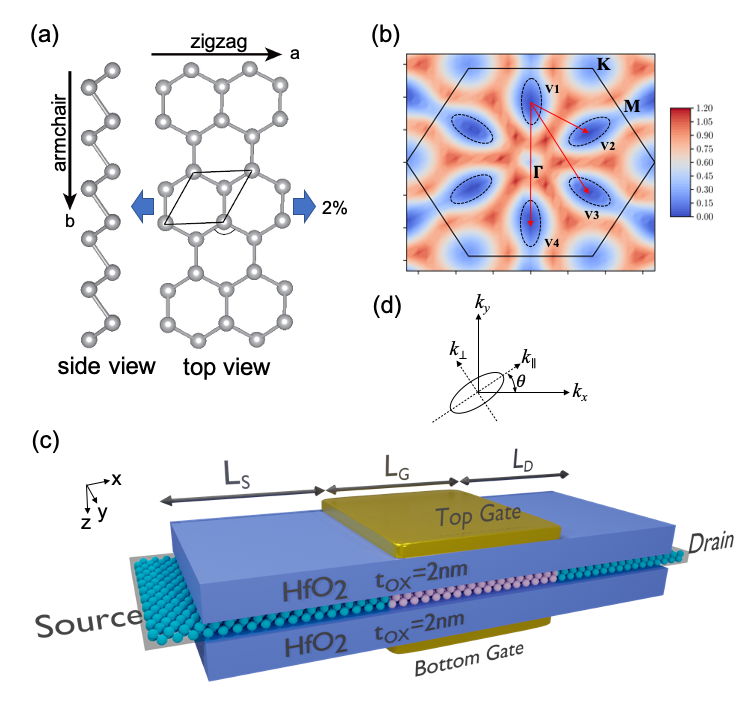}
	\caption{
		(a) Top and side views of 
  the single-layer antimonene and arsenene atomic structure. The unit cell and the direction of the uniaxial strain is also represented. 
		(b) Energy contour plots of the lowest conduction band (CB) in single-layer antimonene. Six-fold degenerate CB valleys are indicated.
		Three types of intervalley scattering channels are represented by red arrows.
		(c) Schematic view of the double-gate field-effect transistor investigated in this work. The  
		gate length $L_G$ varies between 4 and 9~nm. The source and drain extensions are doped with a donor  concentration $N_D$= 
		3.23$\times 10^{13}$~cm$^{-2}$. The doping region is marked by blue color and stops exactly at the channel-source and channel-drain interfaces. Two $t_{\rm OX}$=2~nm
	    thick dielectric layers (HfO$_2$) sandwich the 2D material channel and separate it from the gate electrodes. Transport occurs along the $x$-axis, 
        $z$ is vertical confinement direction, and the device is assumed to be periodic along $y$.
       (d) local $k$-space reference frame ($k_\parallel$, $k_\perp$) of a CB valley and the global one ($k_x$, $k_y$), where $k_x$ and $k_y$ are the wavevector components along the transport and the transverse (periodic) direction, respectively.
}
	\label{fig:sketch-valleys}
\end{figure}

In this paper, we report the results of room-temperature numerical simulations of arsenene- and antimonene-based short-channel MOSFETs with double-gate architecture. Arsenene and antimonene are among the most promising candidates as channel materials in emerging MOSFET devices, due to their small effective masses and correspondingly high theoretical mobility~\cite{wang2017a,pizzi2016a,klinkert2020a}. However, previous application-oriented studies have been based either on purely ballistic simulations or have considered only acoustic phonon scattering. A full  treatment of EPC in these materials is therefore still missing. 

To this purpose, we use a non-equilibrium Green's function (NEGF) transport solver including acoustic and optical intra- and intervalley EPCs with parameters fully determined from first-principles calculations.
The NEGF transport solver is coupled with a Poisson solver for self-consistently computing the electrostatic potential in the whole device domain.

The main findings of this work are not limited to the specific case of arsenene and antimonene, but have more general implications and suggest that a more 
careful treatment of the electron-phonon coupling is necessary  in simulating MOSFETs based on 2D materials in order to avoid severe overestimations of performance.
In view of our results, we {\red investigate the approach proposed by Sohier \emph{et al.}~\cite{sohier2019a}}  to mitigate the detrimental intervalley scattering. This technique is applicable not only to the case of antimonene and arsenene, but also to many similar 2D materials characterized by multivalley transport.

\section{Results and discussion}

\subsection{Conduction band valleys}
\label{ssec:band}
For the purpose of electronic transport simulation, the conduction band of arsenene and antimonene [a single-layer of 
arsenic/antimony atoms forming a buckled hexagonal honeycomb structure, see Fig.~\ref{fig:sketch-valleys}(a)] 
can be described as six energetically degenerate valleys, located {at the midpoint of} the $\Gamma$-M line.
 We note that the next low-energy conduction band valley is 0.23~eV (0.27~eV) above 
 the conduction band minimum for antimonene (arsenene), thus 
 playing negligible roles in the electronic transport at 300~K.
 In our simulations, we restrict ourselves to consider the six low-lying valleys, modeled within an anisotropic effective-mass approximation, as described in Sec.~\ref{ssec:model}.
 We have verified that this is enough to include the relevant states for transport at the doping level and temperature (300~K) considered in this work. 
In this regard, we compare in Fig.~\ref{fig:doublegate-5nm}  the transfer characteristics computed in the ballistic regime within our model for an antimonene-based MOSFET with the corresponding data obtained by using an atomistic full-band model in Ref.~\onlinecite{pizzi2016a}. 
The good agreement  between our results (dashed line) and those obtained within the atomistic full-band model  (dots)  confirms the validity and accuracy of our approach.
{\red The spin-orbit coupling is neglected in this work since it does not affect appreciably the conduction band, as also shown in Ref.~\onlinecite{pizzi2016a}.}

  \begin{table*}
  	\caption{
  		Conduction band effective masses ($m_{\parallel}$ and $m_{\perp}$), 
  		phonon frequencies ($\omega$), and electron-phonon deformation potentials ($\Xi$) calculated from 
  		DFT and DFPT for all the phonon modes in antimonene and arsenene. The values of  $C_{\rm 2D}$  and $\Xi^\Gamma_{\rm AC}$ are taken from
  		Ref~\cite{pizzi2016a}. 
  	}
  	\begin{tabular}{@{}lccc|ccc@{}}
  		\hline
  		& \textbf{antimonene} &  & & \textbf{arsenene}  \\
  		\hline
  		$m_{\parallel}$			  & 0.51     &  &      & 0.46        \\
  		$m_{\perp}$			      & 0.15     &  &      & 0.15        \\
  		$C_{\rm 2D} $ [eV/\AA$^2$]     & 1.948$^\dagger$ && &	3.271$^\dagger$ \\
  		\hline
  		\textbf{Intravalley EPC}          &  &  \\
  		\hline  
  		$\Xi^\Gamma_{\rm LO}$ [eV/cm]    &8.07$\times 10^7$   &  &  &	1.09$\times 10^7$   \\
  		$\Xi^\Gamma_{\rm TO}$ [eV/cm]    &4.90$\times 10^7$  &  &	&  7.03$\times 10^7$   \\
  		$\omega^\Gamma_{\rm LO}$ [meV]                       &20  &  & &	27  \\
  		$\omega^\Gamma_{\rm TO}$ [meV]                       &16	 &  & &   23  \\
  		$\Xi^\Gamma_{\rm AC}$ [eV]    &3.265$^\dagger$   &&	& 3.815$^\dagger$   \\    
  		\hline
  		\textbf{intervalley EPC} & (V1$\rightarrow$V2)         &  (V1$\rightarrow$V3)  & (V1$\rightarrow$V4) &    (V1$\rightarrow$V2)        & (V1$\rightarrow$V3)  & (V1$\rightarrow$V4)  \\
  		\hline  
  		$\Xi_{\rm ZA}$ [ eV/cm]                         & 3.55$\times 10^{7}$   &	9.37$\times 10^{6}$& 	2.22$\times 10^{6}$& 	\textbf{4.78$\times 10^{7}$}	 & 	7.37$\times 10^{6}$	& 	1.74$\times 10^{6}$     \\
  		$\omega_{\rm ZA}$ [meV]                       &   3.50                         &	8.89 &  7.26	                          & 	\textbf{5.81 } &    8.89	                 & 	7.26		    \\    
  		$\Xi_{\rm TA}$ [ eV/cm]                         & 	8.51$\times 10^{4}$ &	2.71$\times 10^{5}$  & 1.26$\times 10^{4}$& 	1.65$\times 10^{6}$   & 	2.16$\times 10^{5}$	& 	1.06$\times 10^{4}$    \\
  		$\omega_{\rm TA}$ [meV]                       &   4.76                          &	8.97	&   8.53                        	& 	7.97	   &    8.97	                 & 	8.53	   \\   
  		$\Xi_{\rm LA}$ [ eV/cm]                         & 	2.841$\times 10^{7}$ &	1.501$\times 10^{6}$   & 1.05$\times 10^{7}$&\textbf{3.97$\times 10^{7}$ }  & 	1.18$\times 10^{6}$	& 	8.24$\times 10^{6}$   \\
  		$\omega_{\rm LA}$ [meV]                       &   6.58                          &	13.90	&   10.07                       	& 	\textbf{9.58  } &     13.89                 	& 	10.07	  \\   
  		$\Xi_{\rm TO1}$ [ eV/cm]                         & 	4.201$\times 10^{5}$ &	\textbf{2.871$\times 10^{8}$}   & 	9.6$\times 10^{3}$& 	9.50$\times 10^{5}$   &   \textbf{2.26$\times 10^{8}$ }	& 	7.53$\times 10^{3}$     \\
  		$\omega_{\rm TO1}$ [meV]                      &   17.0                             &	\textbf{24.08}	 &   24.81                     	& 	24.61	 &   \textbf{24.08	   }              & 	24.81	  \\   
  		$\Xi_{\rm TO2}$ [ eV/cm]                          & \textbf{1.481$\times 10^{8}$ }&	1.961$\times 10^{6}$  & \textbf{1.39$\times 10^{8}$	}&\textbf{1.72$\times 10^{8}$  }&   	1.54$\times 10^{6}$	& \textbf{1.09$\times 10^{8}$  } \\
  		$\omega_{\rm TO2}$ [meV]                      &   \textbf{17.8      }                    &	27.17	   & \textbf{26.19   }              	& 	\textbf{25.85 }&    27.17	                 & 	\textbf{26.19 } \\   
  		$\Xi_{\rm LO}$ [ eV/cm]                            &     \textbf{1.031$\times 10^{8}$} &	2.201$\times 10^{5}$   &   \textbf{9.08$\times 10^{7}$}  &  	\textbf{1.57$\times 10^{8}$  }&    1.72$\times 10^{5}$	& \textbf{7.12$\times 10^{7}$ }\\
  		$\omega_{\rm LO}$ [meV]                        &  \textbf{20.7  }                            &	27.58	 &  \textbf{30.59}           	& \textbf{30.26}    &    27.58	                 & \textbf{30.59} \\   
  		\hline
  	\end{tabular}
  	\label{tab:epc}
  \end{table*}
 
\subsection{Electron-phonon couplings from first-principles}
  \label{ssec:epc}

 We compute the relevant electron-phonon couplings from first-principles by using the density functional perturbation theory (DFPT).
The values of the deformation potentials and phonon mode frequencies of intravalley and intervalley EPCs are listed in Table~\ref{tab:epc}. The details about the DFPT calculations can be found in Sec.~\ref{ssec:dft}.
 As shown in Fig.~\ref{fig:sketch-valleys}(b), three types of intervalley EPCs connecting two electronic states located in different valleys  
 (V1$\leftrightarrow$V2, V1$\leftrightarrow$V3, and V1$\leftrightarrow$V4) are considered. 
 In the transport calculations, all the intravalley EPCs are included, while only the intervalley EPCs with a significant deformation potential ($>4\times10^7$~eV/cm)
 are taken into account. 
Ignoring the smaller EPCs has very little effect on the results, 
since the electron-phonon self-energies depend quadratically on the deformation potential (see Eq.~(\ref{eq:opt}) in Sec.~\ref{ssec:method-transport}). 
{\red The intravalley electron-ZA phonon interactions (not included in Table I) are neglected in our simulations. The validity of this approximation can be checked from Table S I and S II in the Supplementary Information, that report the scattering rates at the conduction band minima associated to each phonon mode. The data show that the intravalley scattering rates between electrons and ZA phonons are extremely small compared to all other scattering processes.}
Recently, it has been shown that in antimonene the electron-transverse acoustic (TA) phonon couplings are forbidden by symmetry for both intra- and intervalley scattering~\cite{wu2021a}, in good agreement with our numerical calculation here. 

From Table~\ref{tab:epc}, we can notice that some intervalley EPCs have deformation potential values 
almost an order of magnitude larger than those of the intravalley EPCs. 
Due to the relatively low phonon frequencies in antimonene and arsenene, those phonon modes can be thermally populated at room temperature,
which leads to increased scattering of electrons through both emission and absorption of phonons. 
Therefore, the intervalley EPCs are expected to play a non-negligible role in the electronic transport and sensibly affect the MOSFET performance.
This is confirmed by our quantum transport simulations, presented in the next section. 

\subsection{Quantum transport simulations with dissipation}
\label{ssec:device}
Our aim is to provide quantitative and precise 
predictions of the optimum performance of these devices, when all the extrinsic sources of scattering are eliminated. 
Under these hypotheses, the electron-phonon interaction represents the dominant scattering mechanism at room temperature.
 
 Based on the electronic model and the EPC models presented in the previous sections, we have built a dissipative 
 quantum transport solver  including all the elastic and inelastic electron-phonon scattering {mechanisms}, 
 while still keeping the computational burden at the minimum level. 
 More details on the quantum transport model are given in Sec.~\ref{ssec:method-transport}.
 
We focus on $n$-type MOSFET devices.
A sketch of the simulated device structure is drawn in Fig.~\ref{fig:sketch-valleys}(c). 
We consider a double-gate MOSFET architecture with source and drain extensions chemically doped 
at a donor concentration $N_D = 3.23\times10^{13}$~cm$^{-2}$
 and gate lengths $L_G$ ranging from 4 to 9~nm. The channel is undoped and has the same length of the gates.
The effective oxide thickness  is chosen to be 0.42~nm (corresponding to a 2~nm thick HfO$_2$ high-$\kappa$
dielectric layer), and the supply voltage is $V_{\rm DD}$=0.55~V, according to the ITRS specifications~\cite{itrs}. 

\begin{figure}
	\centering
	\includegraphics[width=8cm]{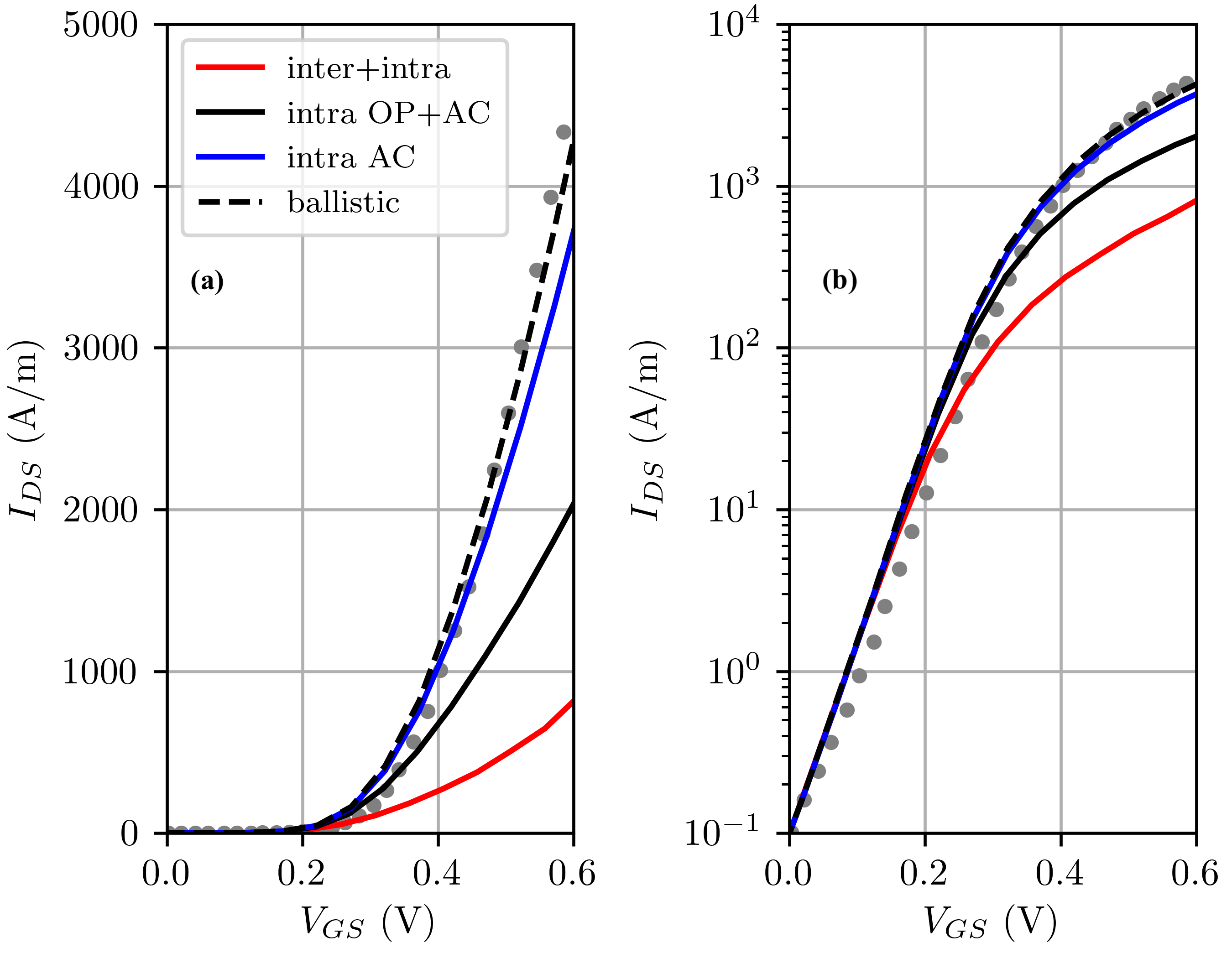}
	\caption{
	$I_{\rm DS}$-$V_{\rm GS}$ transfer characteristic in linear (a) and logarithmic (b) 
scale for an antimonene-based double-gate transistor with $L_G$=5~nm. The considered electron-phonon scattering processes are: intravalley scattering mediated by acoustic phonons (blue line), intravalley scattering mediated by acoustic and optical phonons (black line), and both intervalley and intravalley scattering mediated by acoustic and optical phonons (red line). 
	The results obtained in the ballistic transport regime (dashed lines) are compared to the values (gray dots) reported in Ref.~\onlinecite{pizzi2016a}.  }
	\label{fig:doublegate-5nm}
\end{figure}

We show in Fig.~\ref{fig:doublegate-5nm} the transfer characteristics of 
armchair-antimonene-based MOSFET  with $L_{G} = 5$~nm. 
To illustrate the role played by the different electron-phonon scattering channels
 we carry out the simulations in three cases:  
(1)  including only the intravalley acoustic scattering (elastic), 
(2)  including only the  intravalley scattering (elastic and inelastic), 
and (3) including all the intravalley and intervalley scattering (elastic and inelastic). 

For a clearer comparison of different calculations and  
 following the common practice, we shift all the transfer characteristics 
in order to set the OFF state, chosen to be the device state in which $I_{\rm DS}=I_{\rm OFF}\equiv0.1$~A/m~\cite{itrs}, at $V_{\rm GS}=0$.
Accordingly, the ON state is at $V_{\rm GS}=V_{\rm DD}$=0.55~V, 
and the corresponding current defines the ON current ($I_{\rm ON}$).
The subthreshold swing (SS) is defined as the inverse slope of the $I_{\rm DS}$-$V_{\rm GS}$ curve in 
semi-logarithmic scale in the subthreshold regime ($I_{\rm DS}<1$~A/m), SS=$[\partial \log_{10}(I_{\rm DS})/\partial V_{\rm GS}]^{-1}$.

As shown in Fig.~\ref{fig:doublegate-5nm} for an antimonene-based MOSFET with $L_{\rm G}=5$~nm, the SS is unaffected by the EPC.
However, the ON state performance is strongly degraded, 
especially by the intervalley scatterings, 
consistently with the high {amplitude} of the associated deformation potentials. In quantitative terms, when all the significant scattering processes are included, the ON current drops by a factor larger {\red than} 4 with respect to the ballistic or ADP scattering approximation. Similar results are obtained in the case of the MOSFET based on arsenene and can be found in the Supplementary Information. More in general, these data strongly suggest that modeling the transport in 2D-material-based transistors as elastic, either in the ballistic or in the ADP scattering approximation, is not sufficient to reliably predict the device performance. {On the contrary}, inelastic intravalley and intervalley scattering mechanisms dominate and significantly affect the transport, even at gate lengths as short as 5~nm.

\begin{figure*}
	\centering
	\includegraphics[width=0.47\linewidth]{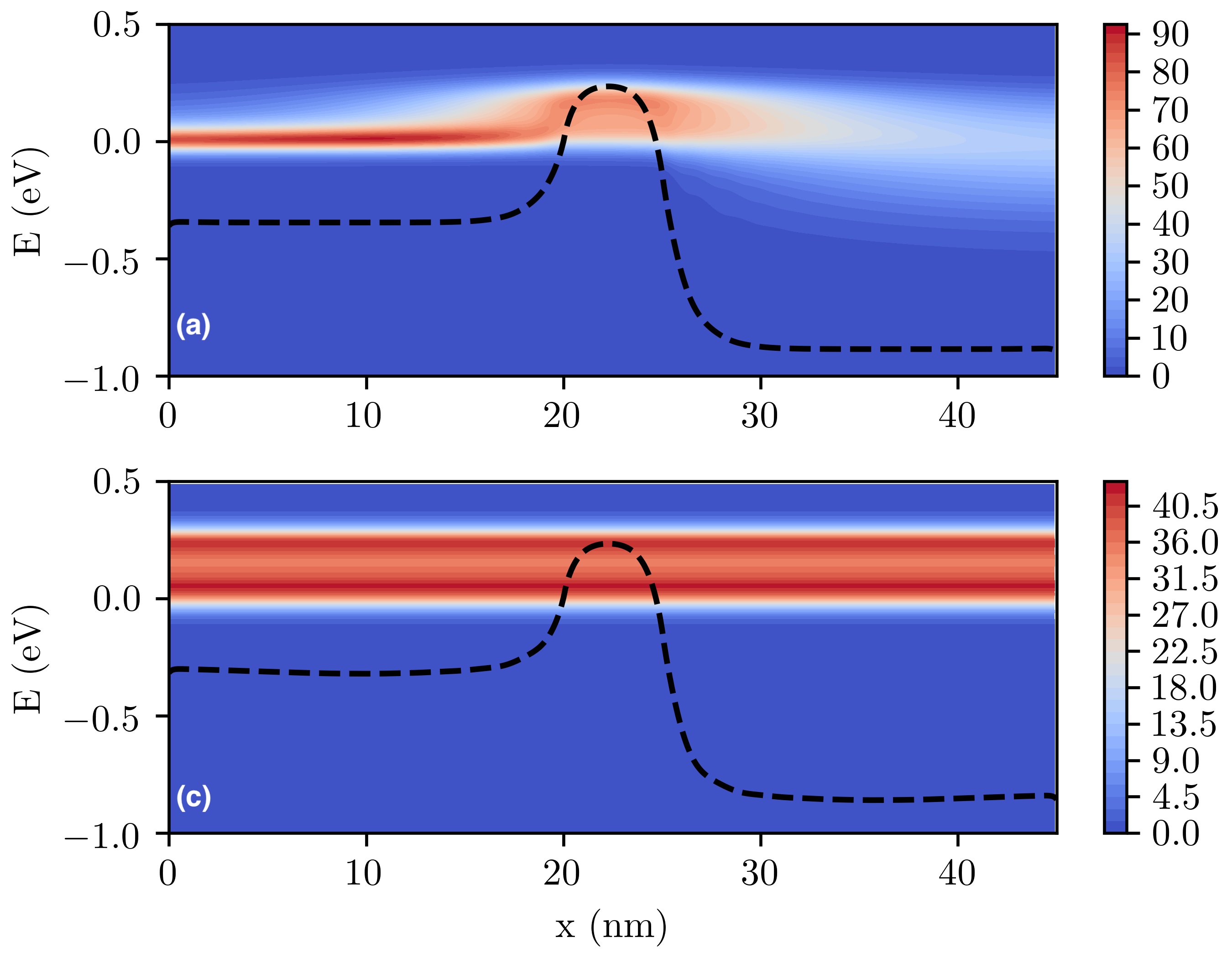}
	\includegraphics[width=0.47\linewidth]{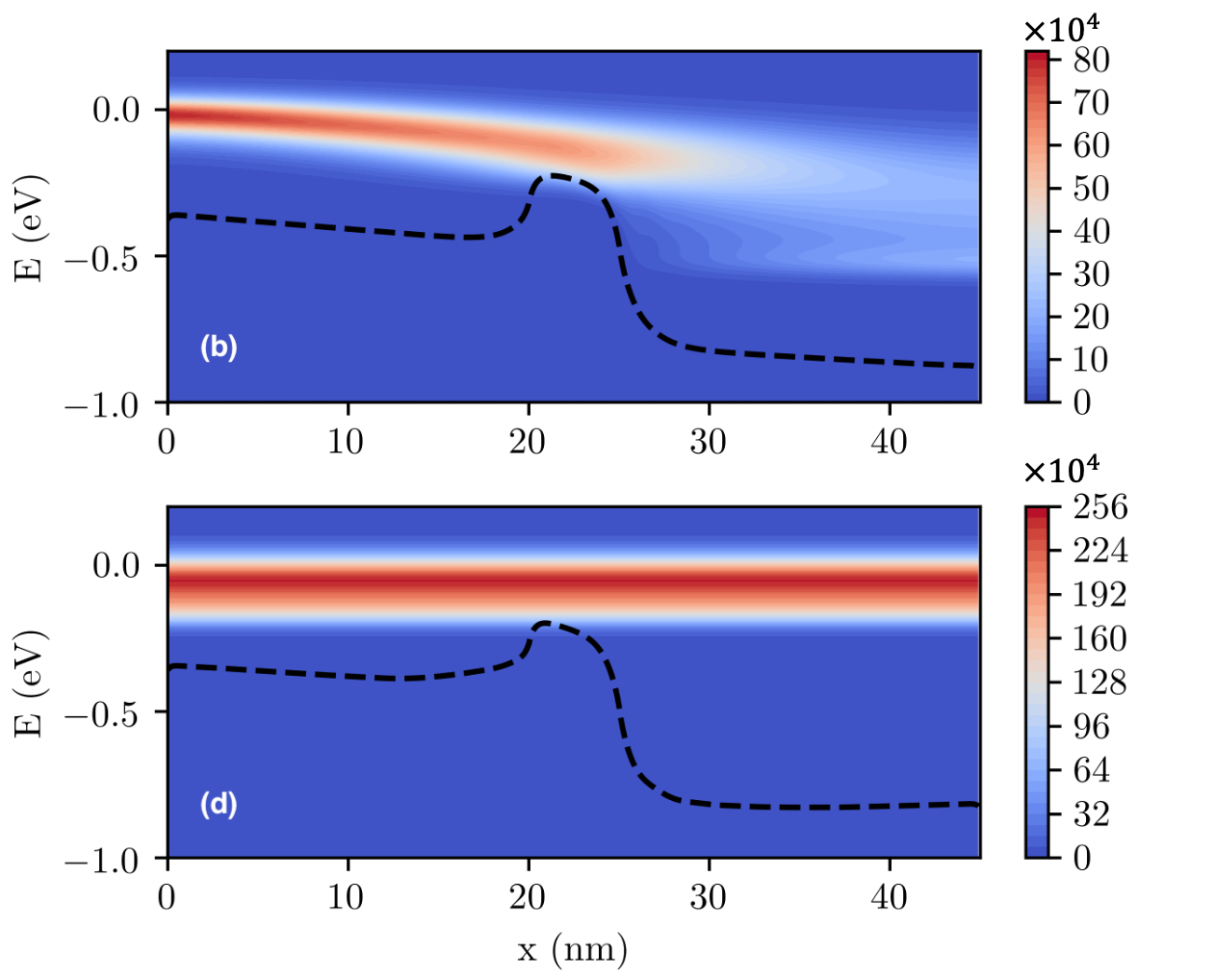}
	\caption{Current density spectrum in an antimonene-based MOSFET with $L_{G} = $5~nm in the OFF state (a,c) and ON state (b,d) considering inter- and intravalley acoustic and optical scattering altogether (a,b), and only considering intravalley elastic acoustic scattering (c,d). The black dashed lines indicate the   conduction band profile. The Fermi level at the source contact is set at 0 eV, while the Fermi level at the drain contact is set at 0.55 eV. The current density spectrum is in arbitrary unit.}
	\label{fig:jx-5nm}
\end{figure*}

To elucidate the effects of EPC on the charge transport all along the device,   
 we plot in Fig.~\ref{fig:jx-5nm} the current spectrum in the OFF and in the ON states for  $L_G = $5~nm. We consider  both the case in which all the relevant scattering processes are included (panels (a) and (b)) and the case in which only the elastic scattering between electrons and acoustic phonons is taken into account (panels (c) and (d)).
We first compare the current spectra in the OFF state. In Fig.~\ref{fig:jx-5nm} (a), it can be seen that the injection energies associated to the inelastic current spectrum gather around the source Fermi level (set to 0 eV). However, close to the barrier the spectrum broadens over a wider energy window, similar to that over which the elastic current concentrates (Fig.~\ref{fig:jx-5nm} (c)). This effect is driven by the significant optical phonon absorption, which rises electrons at energies for which the tunneling probability through the barrier is higher. The net result is that the barycenter of the inelastic current spectrum crossing the barrier is close to that observed in elastic transport conditions.
Moreover, due to the smallness of the charge density in the channel in the subthreshold region, the conduction band profile is unaffected by the transport and only determined by the device geometry and by the dielectric properties of the materials (see the Supplementary Information for related data). Particularly, the efficiency of the gate in modulating the barrier turns out to be independent of the electron-phonon scattering processes taken into account. 
Since the SS is completely determined by the energy distribution of carriers and by the gate modulation efficiency, the previous considerations explain the closeness between the SS values in the elastic and inelastic transport cases. This argument holds respectless of the considered gate length (see the Supplementary Information for the data at $L_G$=9 nm).    

We now focus on the ON state. The current spectrum in Fig.~\ref{fig:jx-5nm} (b) and (d), shows that, contrarily to the behavior in the OFF state, electrons are now mainly injected at higher energies with respect to the top of the channel barrier. The current spectrum in Fig.~\ref{fig:jx-5nm} (b), obtained by including the inelastic scattering processes, demonstrates that electrons gradually lose energy by emitting optical phonons. The strong emission processes entail a significant backscattering and explains the considerable decrease of the current with respect to the case in which only elastic electron-phonon interactions are considered.

\begin{figure*}
	\centering
	\includegraphics[width=0.47\linewidth]{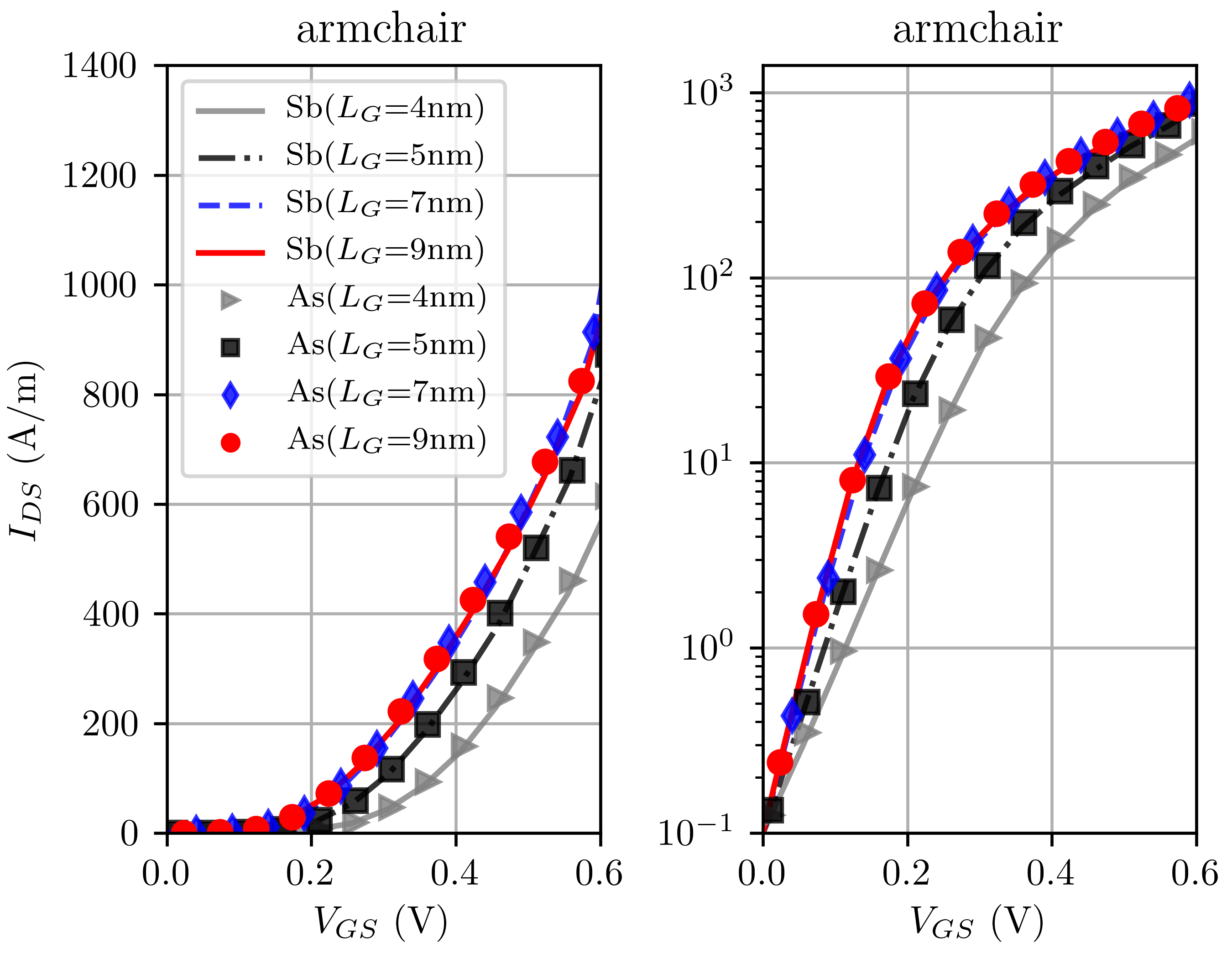}
	\includegraphics[width=0.47\linewidth]{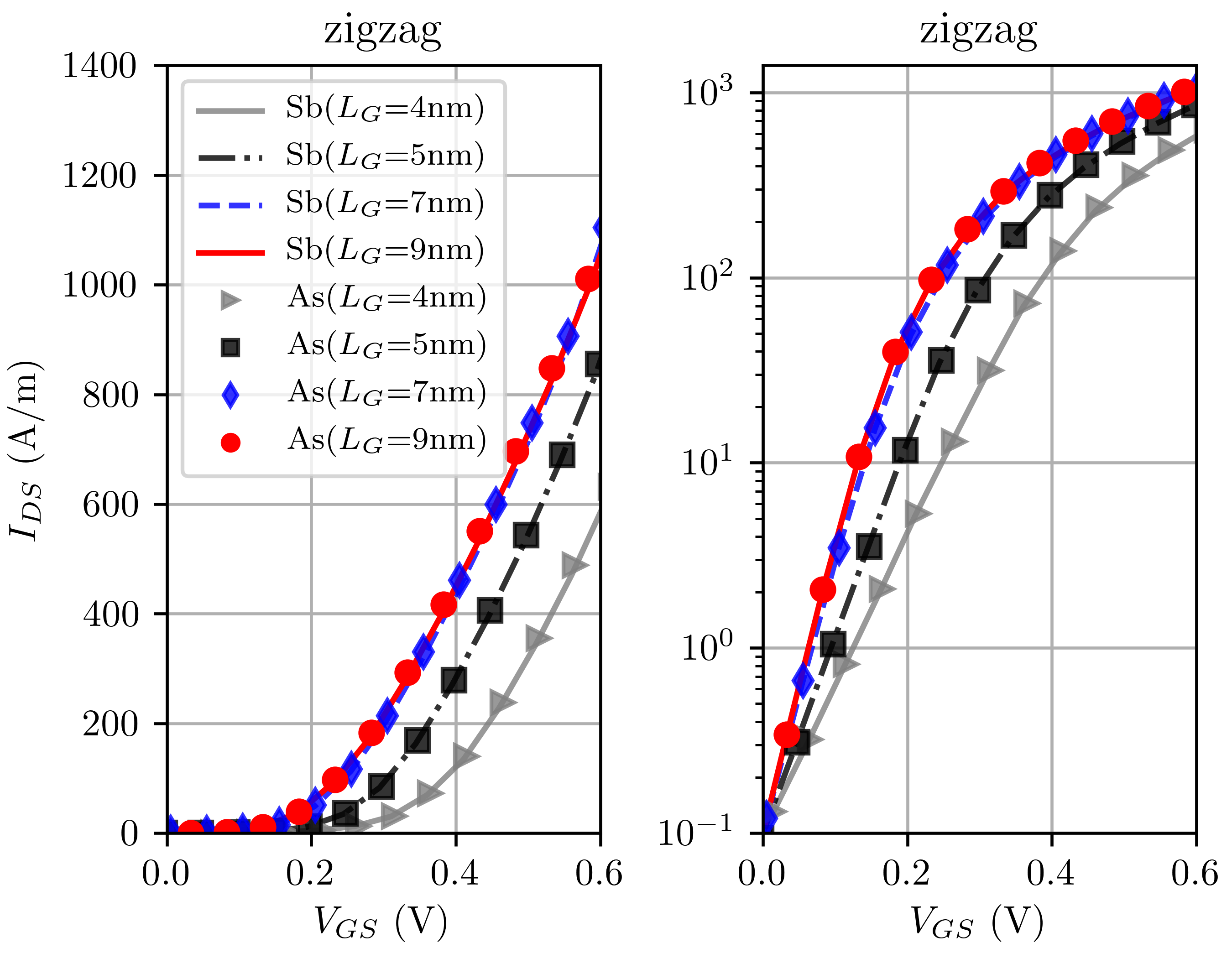}	
	\caption{$I_{\rm DS}$-$V_{\rm GS}$ transfer characteristics in linear and logarithmic 
scale of antimonene (Sb) and arsenene (As) based MOSFETs along armchair and zigzag directions, and for $L_G$=4, 5, 7 and 9 nm.		}
	\label{fig:as-sb-dg-armchair-5-9nm}
\end{figure*}

\begin{figure}
	\centering
	\includegraphics[width=6cm]{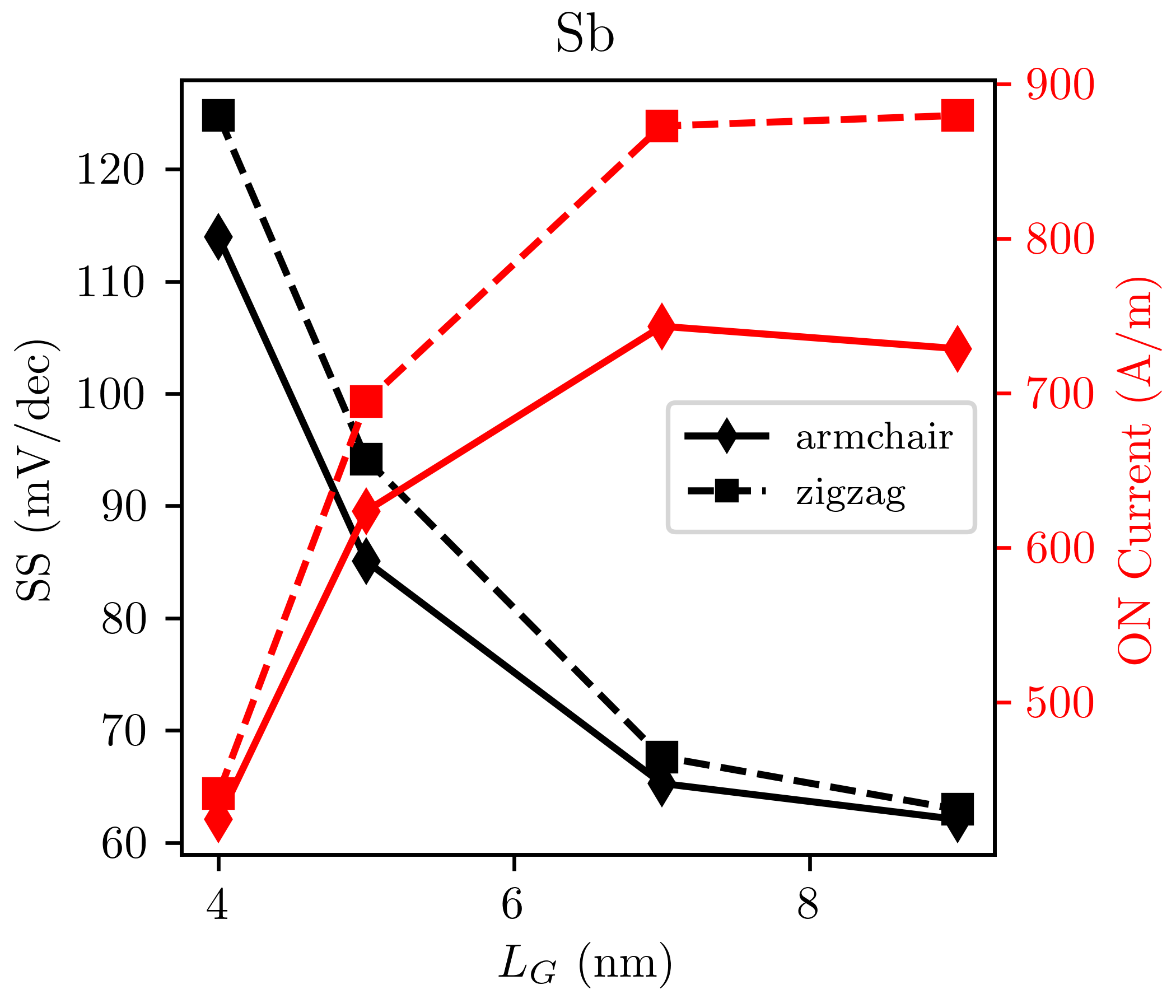}
	\caption{
		ON current and SS as a function of the gate length  $L_{G}$ extracted from the transfer characteristics in Fig.~\ref{fig:as-sb-dg-armchair-5-9nm} of the antimonene-based MOSFET. 
	 }
	\label{fig:ss-ion}
\end{figure}

Figure~\ref{fig:as-sb-dg-armchair-5-9nm} summarizes the simulated  transfer characteristics of the antimonene- and 
arsenene-based MOSFET for $L_G$ varying from 4 to 9 nm, and the transport axis along the armchair
and zigzag directions.
All the relevant electron-phonon scattering processes are included.
 In all the considered cases the antimonene- and arsenene-based devices exhibit very similar transfer characteristics. 
 The extracted SS and the ON current for the antimonene-based MOSFET are reported in Fig.~\ref{fig:ss-ion}. The very similar data obtained for the arsenene-based transistor can be found in the Supplementary Information. We observe that the armchair and zigzag transport directions are not equivalent, since the rotation by $\pi$/4 needed to switch between them is not a symmetry of the crystal. Particularly, while the density of states and the scattering rates associated to the six degenerate valleys are invariant under arbitrary rotations, the anisotropy of the valleys makes the average effective mass seen in the armchair and zigzag directions different. The higher ON current in the zigzag direction arises from a smaller average transport effective mass $\langle m_t^*\rangle$, and thus a higher group velocity. On the other hand, the smaller  $\langle m_t^*\rangle$ entails a larger source-to-drain tunneling (STDT), which results in higher SS values.

\subsection{ Reducing phonon scattering by valley-engineering }
\label{ssec:strain}

\begin{figure*}
	\centering
	\includegraphics[width=0.47\linewidth]{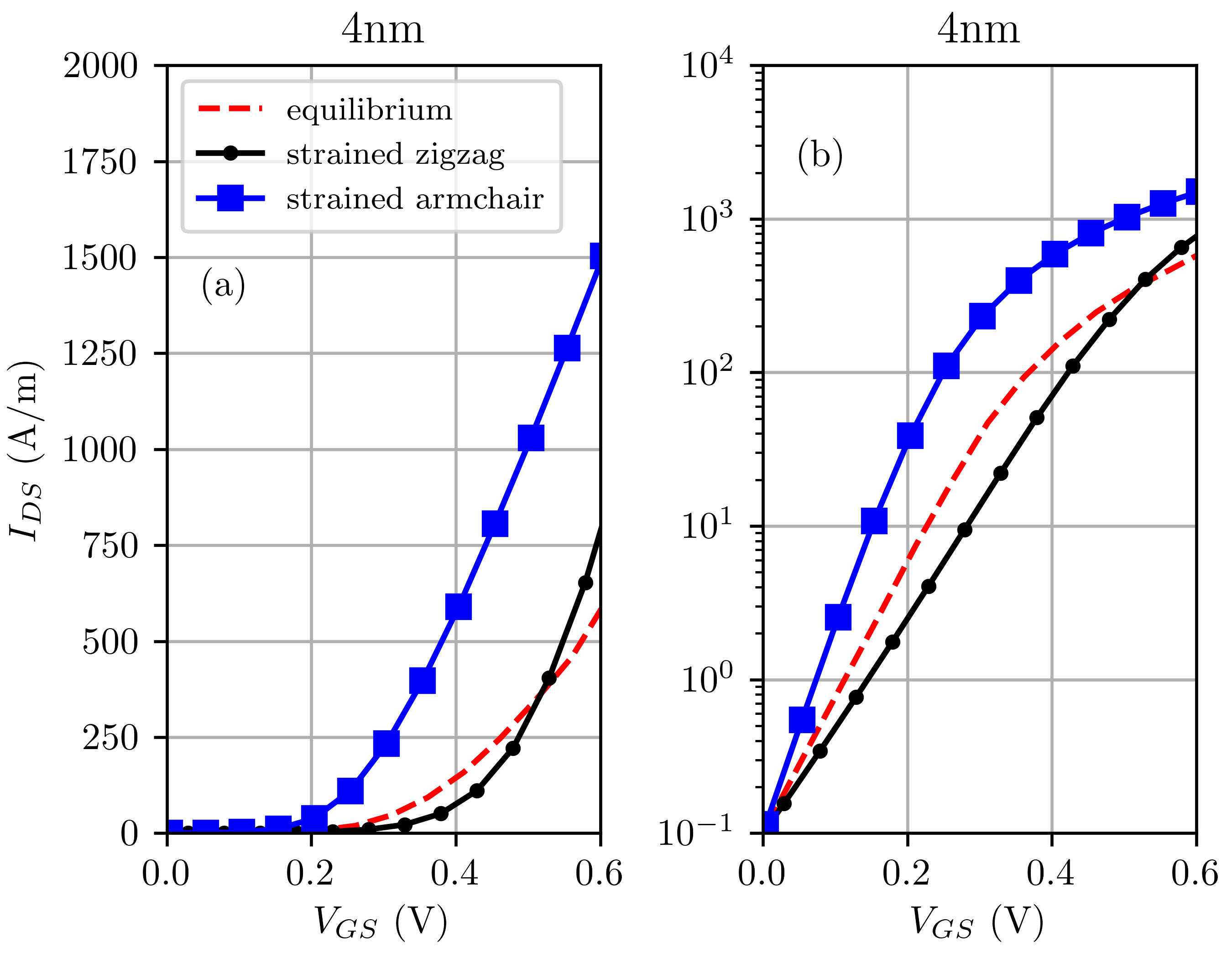}
	\includegraphics[width=0.47\linewidth]{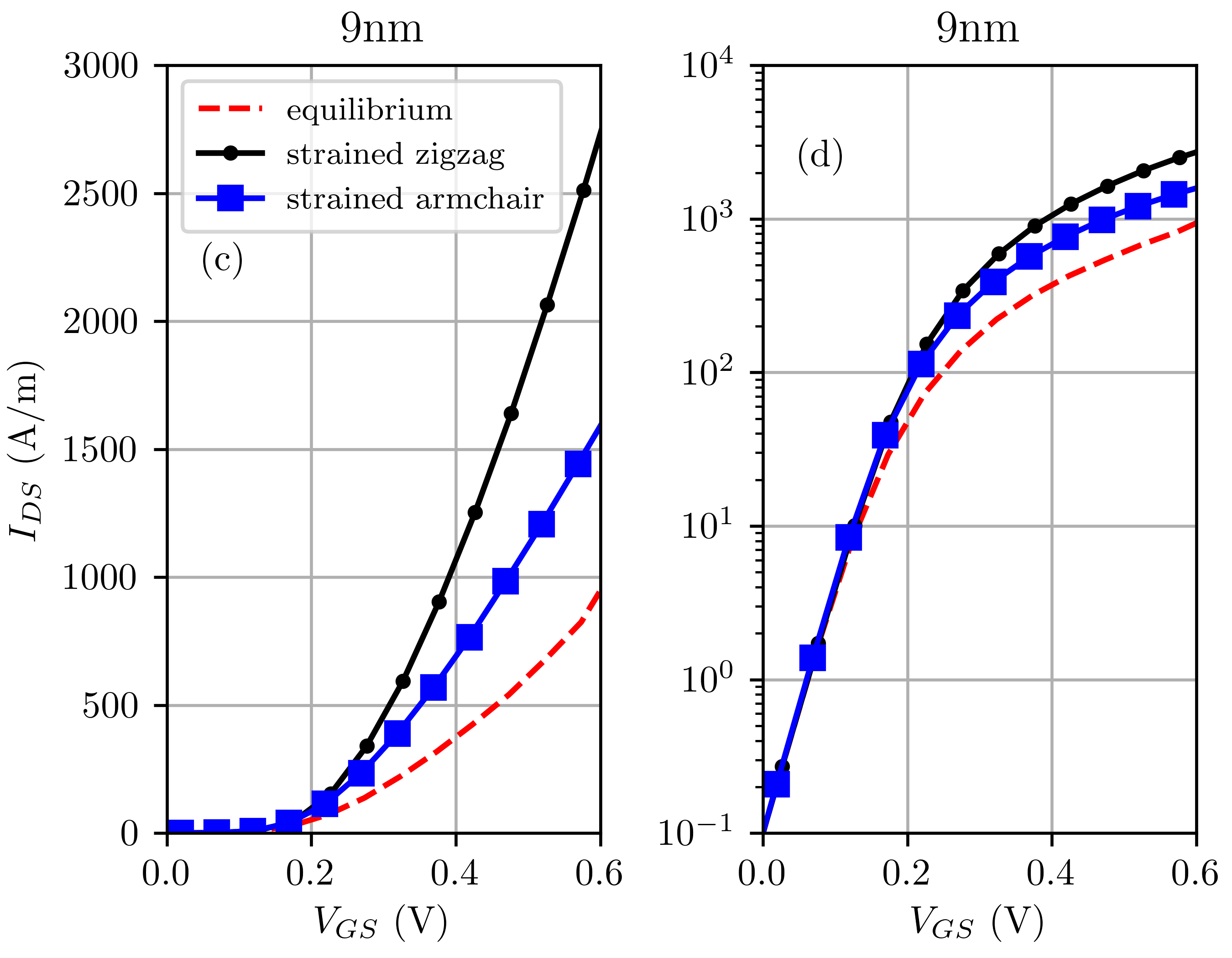}	
	\caption{$I_{\rm DS}$-$V_{\rm GS}$ transfer characteristics in linear and logarithmic scale for a MOSFET based on arsenene with an uniaxial strain of 2\% along the zigzag direction. Two different gate lengths ($L_{G} =$~4 and 9~nm) and transport directions (armchair and zigzag) are considered.}
	\label{fig:strained-4-9nm}
\end{figure*}

\begin{figure}
	\centering
	\includegraphics[width=\linewidth]{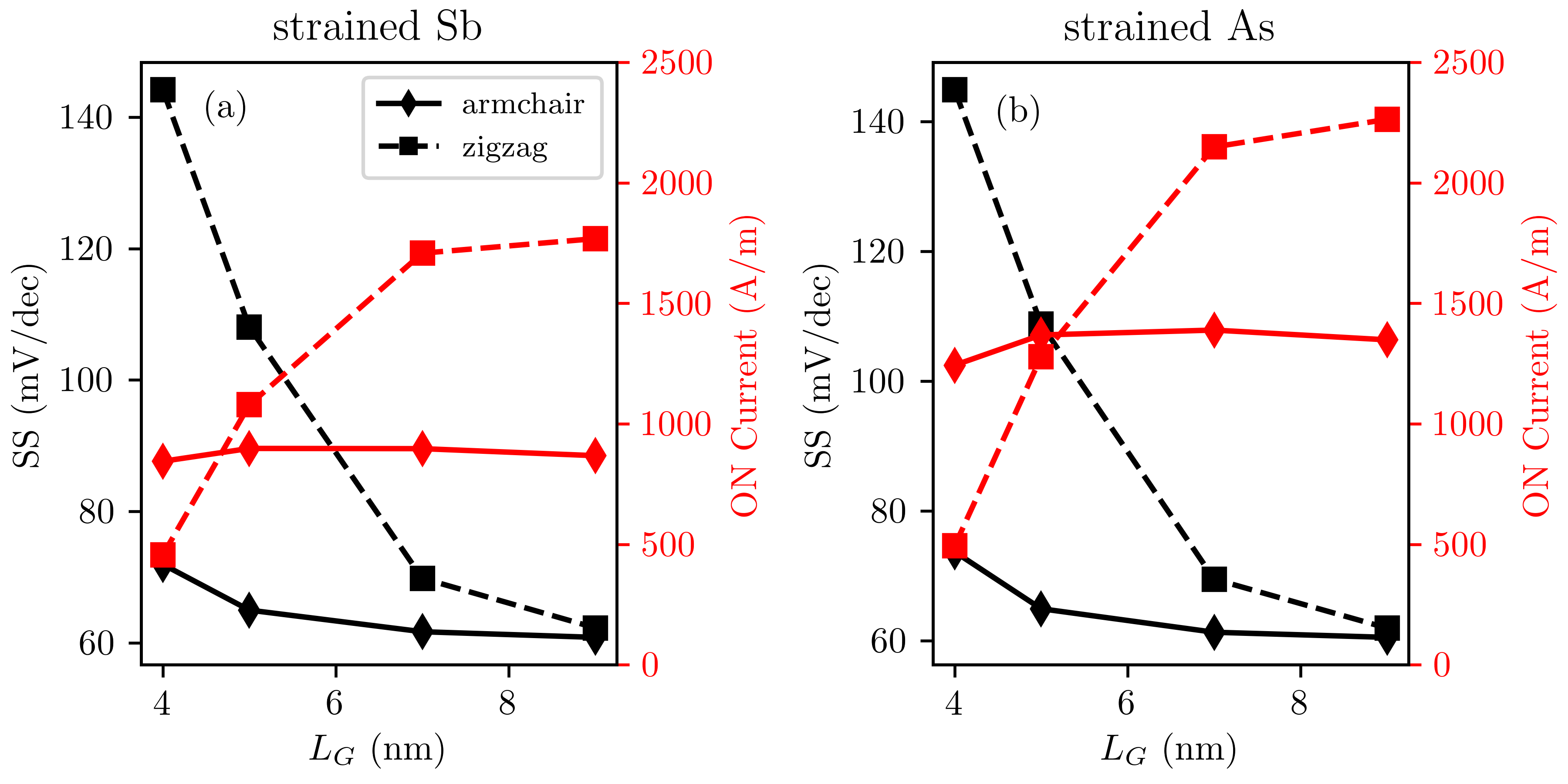}
	\caption{
 ON current and SS as a function of the gate length $L_G$ extracted from the transfer characteristics in Fig.~\ref{fig:strained-4-9nm} of the MOSFETs based on strained antimonene (Sb) and arsenene (As).
 }
	\label{fig:strained-ss-ion}
\end{figure}

Since the six-fold degenerate conduction band valleys are inherited from the crystal rotational symmetry, applying an external
 symmetry-breaking  uniaxial strain along the zigzag direction (see Fig.~\ref{fig:sketch-valleys}) 
 can split the six valleys into four higher-energy  
 and two lower-energy valleys on the armchair axis.
On the contrary, if the uniaxial strain is along the armchair direction,   only two valleys will be pushed up to higher energy~\cite{sohier2019a}.
In the following, we will consider the uniaxial strain along the zigzag direction.
A sizable splitting of $\approx$0.16~eV can be obtained for a small uniaxial strain of 2\% in the case of arsenene, while the splitting is $\approx$0.1~eV in antimonene~\cite{sohier2019a}.
The effect on EPC and effective masses  by such small strain has been shown to be negligible \cite{sohier2019a}.
Instead, since the strain shifts the energy of   the valleys, it effectively turns off the intervalley scattering channels
between the lower-energy   and the higher-energy valleys.
 In addition, the transport becomes anisotropic, since the two lower-lying valleys both have the large effective mass in the armchair direction and the small one in the zigzag direction.

Figure~\ref{fig:strained-4-9nm} compares the transfer characteristics of the strained-arsenene-based MOSFET along the armchair and zigzag directions, for $L_G$=4 and 9 nm. The transfer characteristics of the unstrained device are also plotted for reference. The simulations clearly shows an improvement of the ON current for all the considered configurations. Such an improvement is in excess of a factor 2.5 in the zigzag direction for $L_G$=9 nm and amounts to approximately a factor 2 in the armchair direction for $L_G$=4 nm. Similar improvements (a factor of 2.4 in the zigzag direction for $L_G$=9 nm and ~2 in the armchair direction for $L_G$=4 nm) are obtained for MOSFETs based on strained antimonene. The corresponding data are reported in the Supplementary Information. 

The SS and $I_{\rm ON}$ for both arsenene- and antimonene-based MOSFETs are quantified in Fig.~\ref{fig:strained-ss-ion} for $L_G$=4, 5, 7, and 9~nm. The plots shows that, while the transistors fabricated by using the two monolayers exhibit almost identical SSs as a function of $L_G$, that based on arsenene can achieve higher values of $I_{\rm ON}$. In both cases, at short channel lengths ($L_G$< 5 nm) the best performance are obtained when the transport occurs along the armchair direction. Indeed, due to the higher effective mass along this direction, the $I_{\rm ON}$ is less degraded by the STDT, which results in a better scaling behavior. On the contrary, at longer gate lengths, the transport along the zigzag direction is more advantageous, due to the higher injection velocity associated to the small effective mass. We remark that also the low $m_\perp/m_\parallel$ ratio plays a significant role in enhancing the carrier velocity, since at each given energy it entails a larger kinetic component in the transport direction.

\section{Conclusion}

We have investigated the dissipative transport in single-layer arsenene and antimonene MOSFETs by performing first-principle-based NEGF quantum simulations including all the relevant intra- and intervalley electron-phonon scattering mechanisms.  
We have shown that even for a gate length of 5~nm the ON current can be strongly decreased by the optical intravalley and intervalley phonon scatterings, and that, as a consequence,  the simulations in the ballistic regime and/or only including the acoustic {phonon} scattering can largely overestimate the performance {of real devices}. 
Due to the combination of the six-fold degeneracy of the conduction band valleys, the large magnitude of the electron-phonon couplings, and the low optical phonon frequencies,  
the intervalley electron-phonon scattering is the dominant scattering mechanism in the
arsenene- and antimonene-based MOSFETs at room temperature. 

{Moreover}, we have investigated a viable approach to {mitigate the impact of intervalley phonon scattering} by removing the valley degeneracy via a small uniaxial strain along the zigzag direction.
Our calculations indicate that for gate lengths larger than 5~nm,  
the ON current in the zigzag direction can be increased by a factor 2.5. 
For gate lengths shorter than 5~nm, selecting the armchair direction as the transport direction minimizes the source-to-drain tunneling effect
and results in a two times ON-current improvement with respect to the case with unstrained material.

Overall, our work can provide useful guidelines to the simulation of transistors based on 2D materials, and suggests that there are room and opportunities to overcome the obstacles on the way towards the development of a future 2D-based electronics.

\section{Methods}
\subsection{The anisotropic effective mass model}
\label{ssec:model}

The conduction band of Sb and As monolayers has six-fold degenerate valleys located at the midpoint of the $\Gamma$-M paths. 
These valleys have a larger effective mass along the high-symmetry $\Gamma$-M direction, denoted by $m_\parallel$.  
The effective mass along the direction perpendicular to $\Gamma$-M is denoted by $m_\perp$. 
By using local $k$-coordinates for each valley, we can write the conduction band energy in the anisotropic effective mass model as
\begin{equation}
	\frac{\hbar^2}{2 m_0}\left( \frac{k_\perp^2}{m_\perp} + \frac{k_\parallel^2}{m_\parallel} \right)=E,
\end{equation}
where $\hbar$ is the reduced Planck constant, $m_0$ the electron mass, and $k_\perp$, $k_\parallel$ the wavevector components along the $\Gamma$-M and the orthogonal direction, respectively.
For transport calculations, we need to express $k_\perp$ and $k_\parallel$ in terms of $k_x$ and $k_y$, namely the wave vector components in the transport and transverse directions, respectively (see Fig. 1 (d)): 
\begin{equation}
\left\{ 
\begin{aligned} {}
k_\parallel & = \cos \theta k_x + \sin \theta k_y \\
k_\perp   &  =-\sin \theta k_x+\cos \theta k_y
\end{aligned}
\right. 
\end{equation}
where $\theta$ denotes the angle between the $k_\parallel$-axis and the transport direction $x$. 

The kinetic energy of electrons in terms of $k_x$ and $k_y$ reads:
\begin{equation}
\begin{split}
	\frac{\hbar^2}{2 m_0}&\left[  k_x^2  \left( \frac{\cos\theta^2 }{m_\parallel} + \frac{\sin\theta^2 }{m_\perp} \right)  + k_y^2  \left( \frac{\cos\theta^2 }{m_\perp} +  \frac{\sin\theta^2 }{m_\parallel} \right) \right. \\
	& \left. + 2 k_x k_y \sin \theta \cos \theta \left( \frac{1}{m_\parallel} - \frac{1}{m_\perp} \right) \right]=E. \\ 
\end{split}
\end{equation}
The Hamiltonian in the form employed in the simulations is obtained by replacing  $k_x$  with the differential operator {$-i\partial/\partial x$}. We enforce periodic boundary conditions in the transverse direction, thus the Hamiltonian maintains a parametric dependence on  $k_y$.
The value of $\theta$ to be considered for each valley depends on the chosen transport direction (armchair or zigzag).

\subsection{Quantum transport}
\label{ssec:method-transport}
To simulate the electron transport, we adopt the Keldysh-Green's function formalism \cite{FetterBook}, 
which allows us to include the electron-phonon coupling by
means of
local self-energies. We self-consistently solve the following equations for every valley
\begin{equation}
\label{eq:green}
\left\{ 
\begin{aligned} {}
(EI-H-\Sigma_{\rm SD}-\Sigma_{\rm ph}])  G      =    I  \\
G^\lessgtr  =    G  (\Sigma_{\rm SD}^\lessgtr+\Sigma_{\rm e-ph}^\lessgtr)  G^\dagger  
\end{aligned}
\right. 
\end{equation}
\noindent where $E$ is the electron energy, $I$ is the identity matrix, $H$ is the anisotropic effective mass Hamiltonian matrix,
$\Sigma_{\rm SD}$ and $\Sigma_{\rm e-ph}$ are the retarded self-energies corresponding to source
and drain contacts and to the electron-phonon coupling, $G$ is the retarded Green's function, and 
$\Sigma_{\rm SD}^\lessgtr$, $\Sigma_{\rm e-ph}^\lessgtr$ and $G^\lessgtr$ are the corresponding lesser-than and greater-than quantities.
Equations~(\ref{eq:green}) are spatially discretized over a uniform mesh with stepsize equal to 0.5~\AA.

For the intravalley acoustic phonons, the local self-energies at the $i$-th discrete space site along the transport direction at room temperature can be expressed in the quasi-elastic approximation as \cite{Rogdakis_IOP2009}
\begin{equation}
\label{eq:ac}
\Sigma^{\lessgtr }_{\rm ac, \nu}(i,i;k_y, E)= \frac{\Xi_{\rm ac}^2  k_B  T}{C_{\rm 2D}}  
 \sum_{q_y} G^{\lessgtr }_{\nu}(i,i;k_y - q_y,E),
\end{equation}
where $C_{\rm 2D}$ is the elastic modulus, $\Xi_{\rm ac}$ the acoustic deformation potential, $k_B$ the Boltzmann constant, 
$T$ the temperature, $q_y$ the transverse phonon wavevector  and $\nu$ is the valley index.

For the intravalley optical phonons, the self-energies related to a phonon with branch $\lambda$  
can be expressed as 
\begin{equation}
\label{eq:opt}
\begin{split}
\Sigma^{\lessgtr }_{ \rm \lambda \Gamma, \nu} (i,i;k_y, E) =  \frac{ \hbar \Xi_{\lambda \Gamma}^2}{2 \rho \omega^{\Gamma}_{\lambda }} { \sum_{q_y}}\{  
G^{\lessgtr }_{ \nu}(i,i;k_y - q_y, E \mp \hbar \omega^{\Gamma}_{\lambda }) N(T, \omega^{\Gamma}_{\lambda }) \\ 
+ G^{\lessgtr }_{ \nu}(i,i;k_y - q_y, E \pm \hbar \omega^{\Gamma}_{\lambda }) \left[ N(T, \omega^{\Gamma}_{\lambda }) + 1 \right]   \},
\end{split}
\end{equation}
where  $\rho$ is the mass density, 
 $N(T, \omega^{\Gamma}_{\lambda })$ is the Bose-Einstein distribution at temperature $T$ of a  phonon with frequency $\omega^{\Gamma}_{\lambda }$ at $\Gamma$ point, and 
the upper/lower sign of  the term $\hbar  \omega^{\Gamma}_{\lambda }$ corresponds to lesser/greater-than self-energies. 

The self-energies associated to the intervalley scattering are analogously obtained as:
\begin{equation}
\label{eq:inter}
\begin{split}
\Sigma^{\lessgtr }_{ \lambda \bm{q}, \nu}(i,i;k_y, E) = &  \frac{ \hbar \Xi_{\lambda  \bm{q}}^2}{2 \rho \omega^{ \bm{q}}_{\lambda }}  
\sum_{\nu' \neq \nu} \sum_{q_y}\{  
G^{\lessgtr }_{ \nu'}(i,i;k_y - q_y, E \mp \hbar \omega^{ \bm{q}}_{\lambda })  N(T, \omega^{ \bm{q}}_{\lambda })  \\
& + G^{\lessgtr }_{ \nu'}(i,i;k_y - q_y, E \pm \hbar \omega^{ \bm{q}}_{\lambda }) \left[ N(T, \omega^{ \bm{q}}_{\lambda }) + 1 \right]   \},
\end{split}
\end{equation}
where $\bm{q}$ is the phonon wavevectors connecting two degenerate valley minima, and the three scattering processes  shown in Fig.~\ref{fig:sketch-valleys} are considered.

The total lesser-than and greater-than self-energies  $\Sigma_{\rm e-ph}^{\lessgtr}$  are obtained by summing the self-energies associated to all the different scattering processes. The retarded self-energy is then computed by
\begin{equation}
\label{eq:sig}
	\Sigma_{\rm e-ph}^{ }(i,i;k_y, E) = \frac{1}{2} [	\Sigma_{\rm e-ph}^{>}(i,i;k_y, E) - 	\Sigma_{\rm e-ph}^{<}(i,i;k_y, E) ],
\end{equation}
The real part of the retarded self-energy, which only contributes to a shift of the energy levels, is neglected.

Equations (\ref{eq:green}-\ref{eq:sig}) are iteratively solved within the so-called self-consistent Born approximation (SCBA) until the convergence is achieved. 
The SCBA convergence ensures that the electronic current is conserved 
along the device structure. Once the SCBA loop achieves the convergence, the electron and transport current densities are determined from the Green's functions of all the valleys
\begin{equation}
	\label{eq:rho}
	n(i) = \frac{-i}{2 \pi} \sum\limits_{\nu,\, k_y} \int_{E_{\rm min}}^{E_{\rm max}} \dd{E} G_{\nu}^{< }(i,i;k_y, E), 
\end{equation}
and
\begin{equation}
\label{eq:jx}
\begin{split}
J_x (i) = \frac{e }{h} \sum\limits_{\nu,\, k_y} \int_{E_{\rm min}}^{E_{\rm max}} \dd{E} [H_{i,i+1} G_{\nu}^{< }(i+1,i;k_y, E) \\
 - H_{i+1,i} G_{\nu}^{< }(i,i+1;k_y, E)],
\end{split}
\end{equation}
where 
 the integration range from $E_{\rm min}$ to $E_{\rm max}$  includes all the energy states contributing to transport.
We discretize the energy into a uniform grid with a stepsize $\Delta E=4$ meV for antimonene and $\Delta E=5$~meV for arsenene. 
 Each energy interval of width $\Delta E$ is further discretized by defining 3 quadrature nodes. The charge and current densities are computed by first performing an integration over each interval by using the Gauss-Legendre quadrature rule and then adding the contributions of all the intervals.
The 2D charge density is used to set up the Poisson equation, which is solved in the device $x$-$z$ cross-section.
The simulation runs over a loop between the Green's function equations and the Poisson problem until an overall self-consistent solution is reached.

\subsection{First-principles calculation details }
\label{ssec:dft}

The DFT simulations were performed in a plane wave basis by means of the Quantum Espresso suite~\cite{Giannozzi2009,Giannozzi2017}. The antimonene and arsenene monolayers were simulated by using scalar relativistic norm-conserving pseudopotentials based on the local density approximation exchange-correlation functional. In both cases, a $16\times 16\times 1$ Monkhorst-Pack (M-P) k-point grid was used, with a cutoff energy of  90~Ry. 
A vacuum layer of 28 \r{A} is considered along the perpendicular direction to eliminate the interlayer interactions due to the periodic boundary conditions.
 
The phonon dispersion relations were computed within the DFPT framework by using a 8$\times$8$\times$1 M-P $\bm{q}$-point grid and a 16$\times$16$\times$1 M-P $\bm{k}$-point grid.  

The electron-phonon matrix elements between two electronic states $\ket{n \bm{k}}$ and 
$\ket{n \bm{k}+\bm{q}}$ connected by a phonon mode  $\lambda$ of wavevector $\bm{q}$  
were computed within the DFTP framework as 
\begin{equation}
	g_{\bm{k} \bm{q}, m n \lambda}= \sqrt{\hbar / (2M\omega_{\lambda \bm{q}})} \mel{m \bm{k}+\bm{q}}{\Delta V_{\bm{q} \lambda}^{\rm KS}}{n \bm{k}},
\end{equation}
where $M$ is the unit cell mass, $m$ and $n$ are band indices, and $\Delta V_{\bm{q} \lambda}^{\rm KS}$ is the lattice periodic part of the perturbed Kohn-Sham potential expanded to first order in the atomic displacement.
The $g_{\bm{k} \bm{q}, m n \lambda}$ were computed on a  16$\times$16$\times$1 $\bm{k}$-point grid and a 8$\times$8$\times$1 $\bm{q}$-point grid, and then interpolated on a denser grid by projection on a basis of maximally localized Wannier functions by using the wannier90 code~\cite{wannier90} and the EPW code~\cite{epw}. 
Six Wannier orbitals were employed to reproduce accurately
the conduction and valence bands close to the bandgap and achieve good spatial localization of the Wannier orbitals.
{\red The scattering rates for each phonon mode were computed through the EPW code, by interpolating the electron-phonon matrix elements on a 120$\times$120$\times$1 $\boldsymbol q$-point grid. }
The intravalley and intervalley electron-phonon coupling were described by the matrix element at $\bm{q}=0$  and $\bm{q}=\bm{k'}-\bm{k}$, respectively, where $\bm{k'}$ and $\bm{k}$ correspond to two degenerate valley minima. The deformation potentials were finally computed as 
\begin{equation}
	\Xi_{\lambda q} = |g_{\bm{k} \bm{q}, m n \lambda}| \times \sqrt{2 M \omega_{\lambda \bm{q}} / \hbar}.
\end{equation}

\section{Data Availability}
The data that support the findings of this study are available from the corresponding
author upon reasonable request.

\medskip
\textbf{Supporting Information} \par 

\medskip
\textbf{Acknowledgements} \par 
This work is supported by Natural Science Foundation of Jiangsu Province under Grants No. BK20180456 and BK20180071, and National Natural Science Foundation of China under Grant No. 11374063. 
M.P. acknowledges financial support from the Agence Nationale de la Recherche through the "2D-ON-DEMAND" project (Grant No. ANR-20-CE09-0026-02).
J.C. acknowledges support from the European Unions Horizon 2020 Research and Innovation Programme, under the Marie Skłodowska-Curie Grant Agreement "CAMPVANS" No. 885893.
S.Z. acknowledges financial support from the Training Program of the Major Research Plan of the National Natural Science Foundation of China (Grant No. 91964103).
H.Z. acknowledges financial support from the Shanghai Municipal Natural Science Foundation under Grant No. 19ZR1402900.

\medskip
\textbf{Author Contribution} \par 
J.C., D.L. and M.P. developed the theory and simulation code for the quantum transport calculations. 
J.C. performed the quantum transport simulations and analysed the data.
Y.W. and H.Z. performed the ab-initio calculations. 
M.P., H.Z. and S.Z. designed the simulations.
All the authors discussed the results and contributed to the writing of the manuscript.

\medskip
\textbf{Competing Interests} \par 
The authors declare that there are no competing interests.

\medskip

\bibliography{biblio}

\end{document}